\documentclass[%
    reprint,
    amssymb,
    amsmath,
    aip,
    rsi,
]{revtex4-2}

\usepackage{amssymb}
\usepackage{graphicx}
\usepackage{docs}
\usepackage{bm}
\usepackage[colorlinks=true,linkcolor=blue]{hyperref}
\usepackage{amsmath}
\usepackage{siunitx}
\usepackage{float}
\DeclareSIUnit\amagat{amg}

\usepackage{todonotes}
\usepackage{tabularx, multirow}

\usepackage{color}
\usepackage{ulem}

\usepackage[english]{babel}
\addto\extrasenglish{%
\def\sectionautorefname~#1\null{Sec.\,#1\null}
\def\subsectionautorefname~#1\null{Sec.\,#1\null}
\def\subsubsectionautorefname~#1\null{Sec.\,#1\null}
\def\equationautorefname~#1\null{Eq.\,(#1)\null}
\def\figureautorefname~#1\null{Fig.\,#1\null}
}

\newcolumntype{Y}{>{\centering\arraybackslash}X}

\expandafter\ifx\csname package@font\endcsname\relax\else
\expandafter\expandafter
\expandafter\usepackage
\expandafter\expandafter
\expandafter{\csname package@font\endcsname}
\fi
\def \beq {\begin{equation}}
\def \eeq {\end{equation}}

\begin{document}

\title{Miniature biplanar coils for alkali-metal-vapor magnetometry}  

\author{Michael C. D. Tayler}
\email{michael.tayler@icfo.eu}
\affiliation{ICFO -- Institut de Ci\`encies Fot\`oniques, The Barcelona Institute of Science and Technology, 08860 Castelldefels (Barcelona), Spain}

\author{Kostas Mouloudakis}
\affiliation{ICFO -- Institut de Ci\`encies Fot\`oniques, The Barcelona Institute of Science and Technology, 08860 Castelldefels (Barcelona), Spain}

\author{Rasmus Zetter}
\affiliation{Department of Neuroscience and Biomedical Engineering, Aalto University School of Science, 00076 Aalto, Finland}
\affiliation{MEGIN Oy, 00530 Helsinki, Finland}

\author{Dominic Hunter}
\affiliation{ICFO -- Institut de Ci\`encies Fot\`oniques, The Barcelona Institute of Science and Technology, 08860 Castelldefels (Barcelona), Spain}
\affiliation{Present address: Department of Physics, SUPA, University of Strathclyde, Glasgow G4 0NG, UK}

\author{Vito G. Lucivero}
\affiliation{ICFO -- Institut de Ci\`encies Fot\`oniques, The Barcelona Institute of Science and Technology, 08860 Castelldefels (Barcelona), Spain}

\author{Sven Bodenstedt}
\affiliation{ICFO -- Institut de Ci\`encies Fot\`oniques, The Barcelona Institute of Science and Technology, 08860 Castelldefels (Barcelona), Spain}

\author{Lauri Parkkonen}
\affiliation{Department of Neuroscience and Biomedical Engineering, Aalto University School of Science, 00076 Aalto, Finland}
\affiliation{MEGIN Oy, 00530 Helsinki, Finland}

\author{Morgan W. Mitchell}
\affiliation{ICFO -- Institut de Ci\`encies Fot\`oniques, The Barcelona Institute of Science and Technology, 08860 Castelldefels (Barcelona), Spain}
\affiliation{ICREA -- Instituci\'{o} Catalana de Recerca i Estudis Avan\c{c}ats, 08010 Barcelona, Spain}

\date{August 11\textsuperscript{th} 2022 (revision 2)}

\maketitle

\section*{KEYWORDS}
\noindent Optically pumped atomic magnetometers (OPMs); Biplanar coils; Magnetoencephalography (MEG); Microelectromechanical systems (MEMS). 

\section*{Abstract}
Atomic spin sensors offer precision measurements using compact, microfabricated packages, placing them in a competitive position for both market and research applications.
The performance of these sensors, such as the dynamic range, may be enhanced through magnetic field control.
In this work, we discuss the design of miniature coils for three-dimensional localized field control by direct placement around the sensor, as a flexible and compact alternative to global approaches used previously.  
Coils are designed on biplanar surfaces using a stream-function approach and then fabricated using standard printed-circuit techniques.  Application to a laboratory-scale optically pumped magnetometer of sensitivity approximately \SI{20}{\femto\tesla\per\sqrt\hertz} is shown.
We also demonstrate the performance of a coil set measuring $7 \times 17 \times 17$ \si{\milli\meter\cubed} that is optimized specifically for magnetoencephalography, where multiple sensors are operated in proximity to one another.  Characterization of the field profile using \textsuperscript{87}Rb free-induction spectroscopy and other techniques show $>$96\% field homogeneity over the target volume of a MEMS vapor cell and a compact stray field contour of approximately 1\% at 20 mm from the center of the cell.

\section{Introduction}

Atomic spin devices based on alkali-metal vapors are among the results of the ``second quantum revolution''\cite{MacFarlane2003RSTA361} pushing the boundaries of sensitivity and metrology. They have the potential to reshape technological solutions through ultra-precise clocks for timekeeping \cite{KnappeOL2005}, magnetometers for geomagnetic and biomagnetic investigation \cite{Schwindt2004,budkerromalis} or gyroscopes for navigation and autonomous control \cite{Donley2010,Limes2018}. The miniaturization and mass-manufacturing of atomic devices is also a highly active area of research. So far, miniaturization has been addressed with etched-silicon cells to encapsulate the alkali medium \cite{Liew2004,Karlen2017,Kitching2018} and vertical cavity surface-emitting lasers (VCSELs) for optical pumping \cite{schwindt2004chip}, both of which could be manufactured in large quantities at a competitive cost per unit.

In this paper, we consider the design and manufacturability of another important component of alkali-metal-based sensors: electromagnetic coils. Precisely controlled magnetic fields are central to the proper operation of optically pumped magnetometers (OPMs); in particular, for the variety based on locking to the narrow ground-state Hanle resonance of the alkali-metal atoms around zero field\cite{KnappeNoise,osborne_fully_integrated_2018,Slocum1973}. Due to their simple optics and possibility for compact packaging, a few companies have produced 
commercial prototypes of such OPMs for single- or multi-sensor use \cite{pratt2021kernel,sheng2017microfabricated}. These OPMs are competitive with superconductor-based magnetometers in many applications, at present including magnetoencephalography (MEG)\cite{baillet2017magnetoencephalography}, magnetocardiography\cite{Strand2019}, magnetomyography\cite{Broser2018}, and low-field nuclear magnetic resonance\cite{Savukov2020}, among others.  

Precise control of magnetic fields is required in almost every OPM application. First, as already mentioned, a uniform field may be required to maintain a desired field across the vapor's active volume, whether that be for the Hanle resonance, or for other magnetometers that operate at ambient field\cite{Limes2020,Oelsner2022}. Similarly, the control of field gradients can be a critical factor in applications requiring high magnetic coherence. It is also advantageous to minimize stray fields when two or more sensor units are used in close proximity to one another. In the case of OPMs, a reduced stray field minimizes cross talk and improves sensor dynamic range\cite{Iivanainen2019,Boto2018}, and also ensures a linear, well-defined sensor readout\cite{BORNA2022118818}. Together these requirements define a design goal for magnetic coils, which must be met by mass-manufacturing techniques.

Here we discuss coil sets where electrical current paths are confined to two parallel planes (i.e., biplanar coils) to provide magnetic fields along three orthogonal spatial directions. The biplanar geometry facilitates production using standard printed-circuit-board (PCB) techniques. Each current path is separately optimized via the stream function method\cite{pissanetzky_minimum_1992} to produce a homogeneous central field at the location of the alkali-metal sensor.  As a case study, we demonstrate a miniature biplanar coil set designed to be used with a microelectromechanical systems (MEMS) atomic vapor cell, where the coils must have a stray field below 1\%-2\% at 20 mm distance to minimize cross talk between sensors.  A larger coil set, inexpensively manufactured on a flexible PCB, is also designed to retro-fit an existing laboratory OPM setup.

The paper is organized as follows: \autoref{sec:theory} summarizes design criteria for OPM-motivated biplanar coils using a simple theory of zero-field OPMs and then a coil design work flow based on the stream-function optimization method. Results in \autoref{sec:results} validate the resulting design specification, first by quantifying the magnetic inhomogeneity at the coil center and then by mapping the stray field to deduce the crosstalk performance.  Application to magnetometry is also demonstrated, where coils and the MEMS cell form part of a zero-field OPM with open-loop sensitivity 20 -- 40 \si{\femto\tesla\per\sqrt\hertz}.

\section{Theory}
\label{sec:theory}

\subsection{Near-zero-field OPM response}

Alkali vapors exhibit various nonlinear effects near atomic absorption resonances, such as optical pumping, magneto-optical rotation\cite{RevModPhys.74.1153} and optical transparency\cite{PhysRevA.82.033807} that may be utilized for metrology purposes including magnetometry.  Commercial OPMs produced so far for MEG are based on the ground-state Hanle effect 
, which refers to a magnetic-field-dependent absorption of circularly polarized light by alkali atoms\cite{Happer-Mathur,Shah2007,Castagna2011}.
The characteristics of the effect are: (i) maximum transmission of light through the atomic vapor at zero field $\mathbf{B} =  (B_x,B_y,B_z) = \mathbf{0}$ when atoms are pumped to saturation; and (ii) for light propagation along the $z$ axis, an approximately Lorentzian dependence of transmission versus $B_x$ or $B_y$, which we quantify using the $B_x$ field, giving half-maximum transmission, $B_x=B_1$. 


\newcommand{\supzero}{^{(0)}}

It is common to detect the Hanle resonance using quadrature demodulation\cite{DUPONTROC} of the transmission signal when an oscillating transverse field $\mathbf{B}_{\rm mod}(t) = (B_{\rm mod}\cos(\omega_{\rm mod} t),0,0)$, is applied, with $B_{\rm mod}\sim B_1$. 
The dynamics of the atomic spin polarization vector $\mathbf{P} = (P_x,P_y,P_z)$ under $\mathbf{B}$ and $\mathbf{B}_{\rm mod}$ can be described by the Bloch equations \cite{CohenT-Theory,CohenTannoudji1970}. This is studied by Iivanainen et al.\cite{Iivanainen2019}, from which we derive the results below.  If the total magnetic field is $\mathbf{B}_{\rm total}(t) = \mathbf{B}_{\rm mod}(t) + \mathbf{B}_0 + \mathbf{\delta B}$, where $\mathbf{B}_0$, $|\mathbf{B}_0|\ll B_1$ is the nominal field and $\delta \mathbf{B}$, $|\delta \mathbf{B}| \ll |\mathbf{B}_0|$ is a small deviation to be measured, then defining, for convenience, $\beta_\alpha \equiv B_\alpha / B_1$, the phase quadrature of $P_z$ is

\begin{equation}
\tilde{P}_z
=  P_z\supzero J_0 (p)J_1 (p)  \bm{c} \cdot \delta \bm{\beta} + O(\delta \bm{\beta})^2,
\label{eq:Pz2} 
\end{equation}
where $P_z\supzero$ is the steady-state spin polarization that optical pumping would produce at zero field, $p=|B_{\rm mod}/B_1|$, and $\bm{c}=(c_x,c_y,c_z)$ is the first-order responsivity, scaled such that $\bm{c} = (1,0,0)$ at $\mathbf{B}_0=\mathbf{0}$. The components of $\bm{c}$ are
\begin{subequations}
\label{eq:ccomponentsm}
\begin{align}
c_x &= \frac{1}{1+\beta_x^2} \left(1-J_0^2 (p)\frac{\beta_z^2+\beta_y^2}{1+\beta_x^2}\right), \\
c_y &= \frac{-J_0^2 (p)}{1+\beta_x^2} \left( \beta_z +2 \frac{\beta_x \beta_y}{1+\beta_x^2} \right), \\
c_z &= \frac{-J_0^2 (p)}{1+\beta_x^2} \left(\beta_y +2 \frac{\beta_x \beta_z}{1+\beta_x^2}\right).
\end{align}
\end{subequations}

\begin{figure}
\centering
\includegraphics[width=0.85\columnwidth]{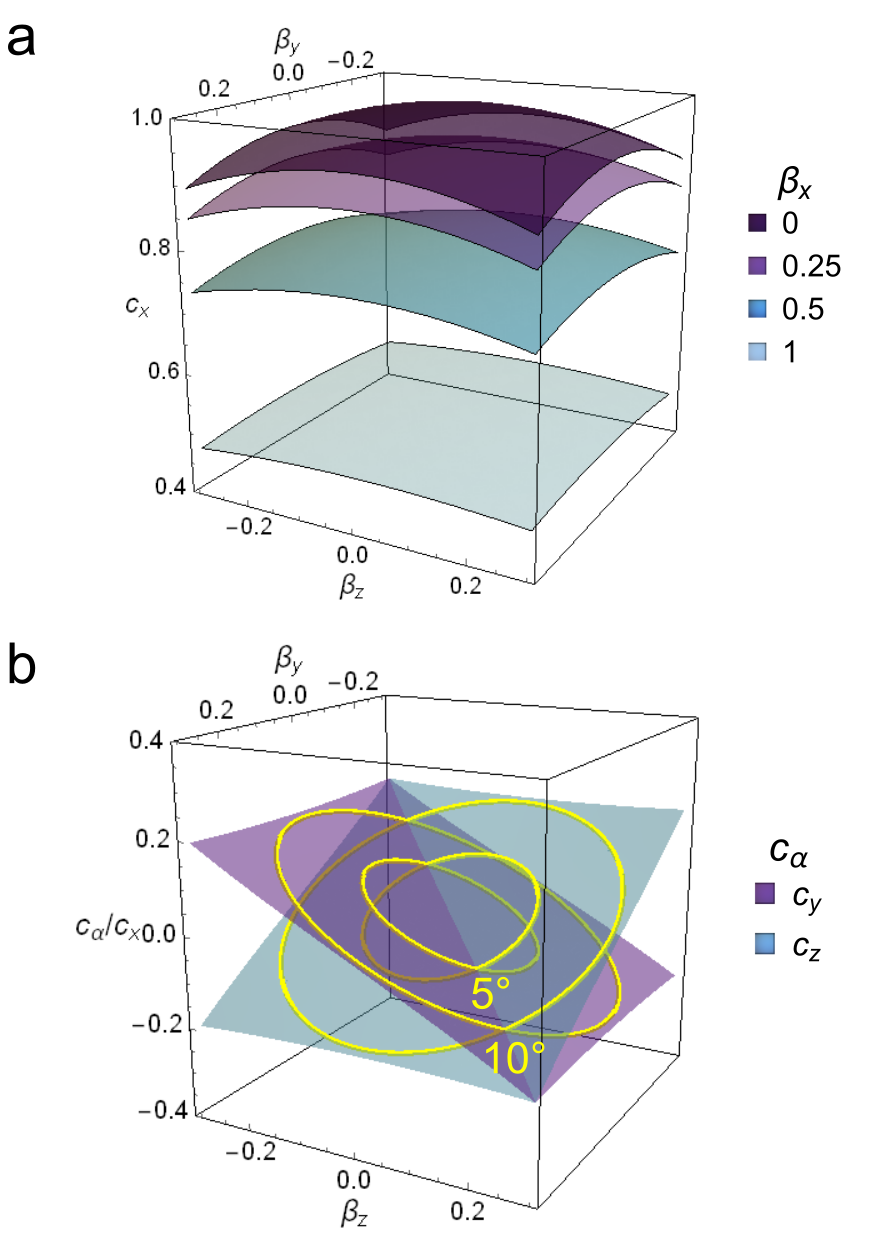}
\caption[Eq3simulation]{The single-beam OPM-scaled responsivity $\bm{c}$ versus the scaled bias field $\bm{\beta}\equiv \bm{B}_0/B_1$  from \autoref{eq:ccomponentsm} with $p=1$: (a) the response coefficient $c_x$ versus $\beta_y$ and $\beta_z$ at fixed $\beta_x$; (b) the sensitive axis perturbations $c_y/c_x$ and $c_z/c_x$ vs.\ $\beta_y, \beta_z$ at fixed $\beta_x=0$. Yellow contours indicate a tilt of $\bm{c}$ away from the $x$ axis. 
}
\label{fig:simulation}
\end{figure}

\autoref{eq:Pz2} is convenient because it allows one to separate $\delta\bm{\beta}$, the signal that we want to measure, from $\bm{\beta}$, the background field.  First, the amplitude of the quadrature signal is highest for $\beta_x = 0$, thus motivating operation as close to zero field as possible.  As shown in \autoref{fig:simulation}, a field bias $\mathbf{B}_0 \neq \mathbf{0}$  produces various changes in the responsivity, relative to the ideal zero-field condition.  The on-axis component $\beta_x$ reduces the desired response $c_x$.  The off-axis components $\beta_y$ and $\beta_z$ produce both a reduction of $c_x$ and an increase in the undesired components $c_y$ and $c_z$.  Such distortions, both in the magnitude and direction of the responsivity, can have a significant impact on the quality of MEG reconstruction\cite{nurminen_effects_2008}.

The spatial homogeneity of $\mathbf{B}$ is also important. Most immediately, inhomogeneity of $\mathbf{B}_0$ within the active volume implies that different locations experience different responsivities $\mathbf{c}$. Due to the negative curvature of $c_x$ with respect to each component of $\bm{\beta}$, inhomogeneity generically implies a reduction in the volume-averaged $c_x$ relative to $c_x$ at the centre of the active volume.  The magnitude of this effect can be roughly estimated by comparing the gradient to  $B_1$ divided by the linear dimension of the active volume,  which is typically of the order \SI{10}{\nano\tesla\per\milli\meter}. Second, since OPMs are densely packed in the MEG application, the field produced by one OPM at the location of its neighbours is of natural concern. In this context, \autoref{eq:ccomponentsm} can be used to find limits on acceptable stray fields. 

Together, these zero-field conditions can be used to define constraints on the components of $\bm{\beta}$ for an OPM design.

\subsection{Field control for zero-field OPMs used in MEG}

In MEG, a few to a hundred zero-field OPM sensor modules are mounted in an array around a subject's head, with the aim of mapping the magnetic field normal to the scalp surface.  Dipole source localization based on inverse modeling of the surface field map requires the output of each OPM to be proportional to the normal field component\cite{nurminen_effects_2008, tierney_spherical_2021}, although a small time-independent component at each sensor is sometimes allowed, overall motivating the magnetic field over the entire array to be nulled as far as possible.  At present, magnetically shielded rooms\cite{hill_multi-channel_2020, pratt2021kernel} or smaller person-sized magnetic shields\cite{Borna_2017,borna_non-invasive_2020} are used to passively attenuate background magnetic fields down to tens of nanotesla and then shim coils inside the shield may further cancel residual fields at the head position into a range where arrayed OPMs are operable.  

Active field-control techniques have more recently shown the potential for great increase in the usability of OPM-based MEG.  Also known as closed-loop OPM operation, these techniques calculate from the demodulation signal a compensatory $x$ field that is fed back to the sensor in real time, giving $\beta_x=0$ to ensure a proportional readout and also increase sensor dynamic range.  Field-feedback loops using large coils to compensate for drift in background fields over the entire sensor array have demonstrated improved operation in suboptimal shielding\cite{holmes_balanced_2019,pratt2021kernel,Iivanainen2019,Kutschka2021} and improved ``wearability'' by ensuring linear signal output from OPM sensors while a subject's head is in motion\cite{boto2018moving}.  The global control approaches demonstrated thus far using large, single coils are, however, limited in temporal bandwidth and cannot fully null field gradients over the array\cite{Packer2021PRA15}.  

A more comprehensive feedback solution with potential to further extend the possible range of head movements or tolerate worse shielding and/or gradients is to locally null the field at each OPM sensor using a miniature set of coils.  The main challenge of localized field control is to operate multiple field-feedback loops without introducing crosstalk.  Here, the source of crosstalk is where a time-dependent compensatory field at one OPM sensor is non-compensatory at its neighbors and may significantly alter the measured magnetic field component \cite{BORNA2022118818}.  At this point, the stray fields produced between neighboring OPMs become of concern and indeed may be considered the dominant source of error in the response.  One strategy to mitigate cross talk would be to limit stray fields from the local coils, so that these are small compared to $\delta\beta_x$; for instance, between two OPM sensors $j$ and $k$, imposing a limit $|a_{k,x}^{(j)}\beta_x^{(k)}|\ll|\delta\beta_x^{(j)}|$ where $\delta\beta_x^{(j)}$ is the target field measured at $j$, $\beta_x^{(k)}$ is the feedback field at $k$, and $a_{k,x}^{(j)}$ is the fraction of $\beta_x^{(k)}$ seen at $j$.   A reasonable limit based on values $\delta B_x \sim 10$ pT, a typical line width of $B_1\sim 10$ nT, and the assumption that the OPMs are all modulated in phase would be $a < 0.01$ between adjacent sensors in the array.

\subsection{Coil optimization via stream function method}

In principle, OPM crosstalk can be completely avoided by suitable coil design.  Magnetostatics allows a current distribution to produce a uniform field with zero gradient and zero fringing field; for example, the so-called ``fluxball'' current geometry\cite{Everett_1966_fluxball}. In practice, good approximations to this ideal can be produced, even given geometrical constraints such as confinement of the currents to application-defined surfaces\cite{makinen2020magnetic} (e.g.\ printed circuits).  

In this work, we employ a stream-function method\cite{pissanetzky_minimum_1992} to find current densities that best satisfy the constraints imposed. Specifically we use the implementation described by Zetter et al.\cite{makinen2020magnetic, zetter2020magnetic} to find line-current paths on planar two-dimensional surfaces for PCB fabrication.
The first step of the method is to define the surface regions upon which currents can be placed.  Constraints related to the surface current are also defined, such as components of the magnetic field at positions inside and outside the surface or constraints related to eddy currents as well as high-permeability shielding in the vicinity of the coil structure.  The next step involves discretizing a current density function on the surface using a triangle mesh and then, finally, numerically optimizing the current density against an objective function to find the best discrete current pattern fulfilling the constraints.  Typically, the objective is to minimize either the inductance or the resistance of the surface current.

\section{Methods}
\label{sec:methods}

\begin{figure*}[t]
\centering
\includegraphics[width=\textwidth]{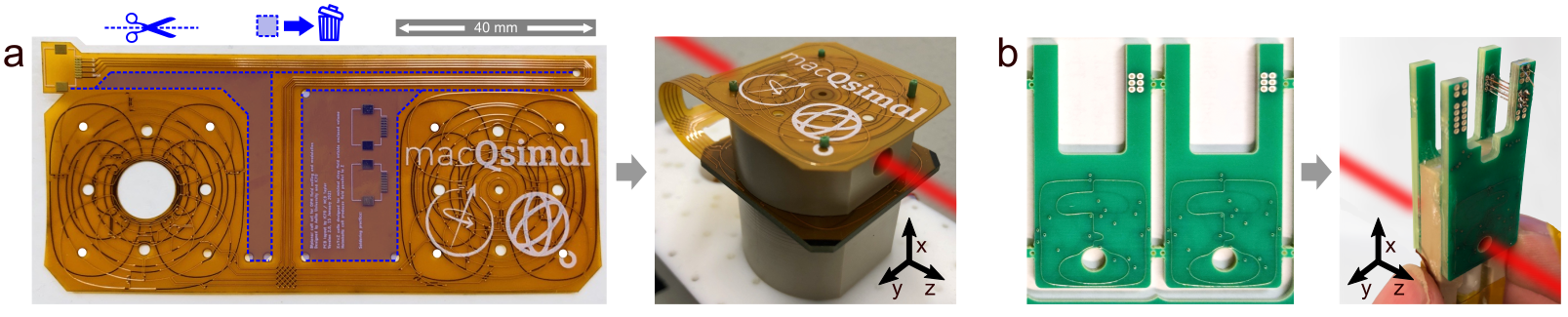}
\caption{Photographs of miniature PCB coil sets as received from the manufacturer and when integrated into the OPM oven setup; (a) A flexible two-layer PCB biplanar coil comprising the three high-homogeneity coils (X, Y, Z) and a Helmholtz coil along the $x$ direction (XH).  The small circular holes in the board are used for alignment and mounting, while the large hole accommodates the OPM heater element.  The distance between the two coil planes is 15 mm. (b) Multilayer FR4 PCB with interplane distance of 6.3 mm, comprising three coils (X, Y, Z) optimized for minimal stray field.}
\label{fig:coil photographs}
\end{figure*}

\subsection{Coil design and fabrication}

\subsubsection{Oversize coils}
Continuous surface-current densities on biplanar surfaces are separately optimized for each X, Y and Z coil using the ``bfieldtools'' software package\cite{zetter2020magnetic, makinen2020magnetic}.  The field specification used in the optimization of each coil consists of a target-field specification and a stray-field specification.  For the target field, a homogeneous field with $<$0.25\% error per field component is specified within a spherical volume of \SI{2}{\milli\meter} radius located halfway between coil planes. The region is discretized on a Cartesian grid with 12 points along each axis, discarding points further than \SI{2}{\milli\meter} away from the grid center to attain the spherical shape. For the stray-field specification, less than 2\% of the target field (for each field component) is allowed at the surface of a cylinder with an inner diameter of \SI{140}{\milli\meter} and a length of \SI{160}{\milli\meter}. The cylinder is also centered on the grid and represents the internal dimensions of the magnetic shield in which the OPM oven is to be mounted.  In practice, this stray field constraint has minimal effect on the design.

Continuous surface current densities on the coil planes are optimized using a minimum-inductance objective.  For each coil, the optimized surface current density is then discretized into individual current loops by extracting a number of equally spaced isosurfaces from the scalar stream function. The number of isosurfaces directly translates into the number of coil windings; a larger number means that the discretized current better approximates the continuous current density and also yields a higher field-to-current ratio, but also results in tighter spacing between loops. 
The isosurface number is chosen heuristically by weighing the desired field parameters against manufacturability. For the Y and Z coils, the number is 6, while for the X coil the number is 8.

The set of coils is optimized initially using inter-plane distances of a four-layer 1.6-mm-thickness FR4 fiberglass PCB; the design is changed mid-process, however, to an all-in-one flexible two-layer PCB.  The flex design is chosen to facilitate installation on the OPM oven setup by avoiding the need to manually install connection wires between the planes, and also because of lower fabrication costs. 

Current paths for each coil are then assigned to top or bottom layers of a two-layer PCB as 0.15-mm-wide traces using a standard PCB-layout editor software program (KiCAD).  Connecting wires between each coil loop are manually routed and extra out or return wire paths are kept spatially close together to maintain faithfulness to the intended magnetic field profile to the maximum extent.  Connections between layers (vias) and breakout solder pads are also manually placed. The lower layer of the $x$ coil is manually modified by adjusting the radius of the innermost current loop to accommodate manufacturing of all four coils on a two-layer PCB.  This post-hoc modification is expected to slightly affect the field-to-current ratio without significantly affecting the field homogeneity.  A Helmholtz coil (XH) with principal axis along the $x$ direction is also routed on the two-layer PCB.

Exported design files (gerber format) are then sent to a PCB fabricator (PCBway, Shenzhen, China), for production of the coils on a \SI{0.2}{\milli\meter}-thick flexible polyimide substrate (\autoref{fig:coil photographs}(a)).

\subsubsection{Miniature coils}
The miniature coil system is designed for a compact, mass-manufacturable OPM.  The biplanar coils as illustrated in \autoref{fig:coil photographs}(b) comprise two stacks of square planes on either side of a cylindrical target region corresponding to the intended position of a MEMS atomic vapor cell, of diameter 3.1 mm and thickness 1.5 mm.  The upper and lower coil plane stacks are +3.5 mm and $-$2.8 mm from the target region centre, respectively, and have a size of \SI{17}{mm} by \SI{17}{mm}.  The symmetry axis of the cylinder is located at a distance of 4.5 mm from the edge of the coil surfaces to reduce sensor standoff distance in the target OPM.

Current densities are optimized under a stray-field constraint specified as 2\% on a cylindrical surface grid centered on the target region, of length \SI{30}{mm} and radius \SI{20}{mm}. Unlike the oversize coils, the stream-function optimization is performed using a minimum-resistance quadratic objective in order to maximize the smoothness of the current patterns.  Furthermore, the discretization from continuous surface-current density to discrete current isosurfaces is manually optimized to ensure best balance between performance and manufacturability; instead of choosing an integer $N$ of equally spaced isosurfaces, discretized coil candidates are generated by choosing every $L_{\mathrm{th}}$ isosurface from a set of $LN$ equispaced isosurfaces. Coil candidates are generated for $N$ = [6, 8, 10] and $L$ = [1,...,16]. Finally, the candidate with the best combination of homogeneity and stray field is chosen, subject to minimum trace spacing of \SI{0.15}{\milli\meter}.

Like the oversize coils, current paths for each coil are then imported into a PCB layout editing software (KiCAD) and assigned to different layers of a multilayer PCB.  Connecting wires between each coil loop, connections between wires, and breakout solder pads are manually routed.  The design is finally manufactured on a standard FR4 PCB substrate (Multi Leiterplatten GmbH, Brunnthal-Hofolding, Germany).

\subsection{Test setup components}

\subsubsection{Atomic cell and oven}
The biplanar coils are tested in a laboratory environment as part of a zero-field-resonance OPM.  A 1.5-mm-thick MEMS vapor cell containing \textsuperscript{87}Rb and N\textsubscript{2} buffer gas (CSEM, Neuch{\^a}tel, Switzerland) is placed in a slotlike cavity inside a ceramic oven (Shapal Hi-M soft) of diameter 15 mm and height 15 mm.  The flexible biplanar PCB coil set is attached to the top and bottom faces of the oven.  Protruding from the base of the oven along the $x$ axis is a cylindrical ceramic pillar of height 20 mm and diameter 8 mm.  The pillar is wrapped with nichrome wire to provide resistive ac heating: a mean electrical power of 10 W at $\sim30$ \si{\kilo\hertz} heats the oven to a temperature above \SI{150}{\celsius}.  A hole is also made through the top part of the oven along $z$ to allow the laser beam to pass.  This assembly is contained inside a polyetheretherketone (PEEK) structure as illustrated in \autoref{fig:coil photographs}(a), for mechanical support and additional thermal insulation.  

\subsubsection{Magnetic shield}
All measurements using the atomic vapor cell, cell oven and biplanar coils are made inside a four-layer mu-metal shield (Magnetic Shields Ltd, UK). Thermal magnetic noise of the shield is approximately \SI{12}{\femto\tesla\per\sqrt\hertz} at 20 Hz and the residual dc field along each axis is below \SI{50}{\nano\tesla}.  No other coils, except for the local biplanar coils placed around the oven, are present inside the magnetic shield.

\subsection{Free-induction-decay measurements}
The magnetic field-to-current ratio and field homogeneity generated across the target region of each biplanar coil are determined by measuring free-precession of optically polarized $^{87}$Rb in the MEMS vapor cell. 

A single elliptically polarized beam of wavelength near to the \textsuperscript{87}Rb D\textsubscript{1} line is used to optically pump and probe the atoms\cite{Shah2009}.  The beam is sourced from a DBR laser located outside the magnetic shield, while the coils and oven assembly are placed inside the shield.  The pulse sequence shown in \autoref{fig:FID signals}(a) shows the temporally separated pump and probe stages of the measurement. During optical pumping, a magnetic field of $\bm{B}_1 \approx (300,0,0)$ \si{\nano\tesla} is applied and the laser frequency is modulated at the Larmor frequency to resonantly drive atomic coherence \cite{grujic2015sensitive, hunter2018free, hunter2018waveform}, producing a component of atomic polarization orthogonal to the beam propagation axis.  The magnetic field is then rapidly switched to a field $\bm{B}_0$, the major component of which is along one of the $x$, $y$, or $z$ axes, and the free precession is probed via the time-dependent optical rotation induced in the linearly polarized component of the laser beam.  A balanced photodetector (Thorlabs PDB450A) connected to a digital acquisition card (NI-DAQ PCI-4462) is used to record the intensity of \textit{s} and \textit{p} polarization components.

For each coil the linear slope of the free-induction-decay (FID) frequency is measured versus the applied current and divided by the atomic gyromagnetic ratio ($\gamma/(2\pi) = 7.0$ \si{\hertz\per\nano\tesla}) to yield the field-to-current ratio. 

\subsection{Stray field mapping}

Stray magnetic field profiles are measured with a commercial, self-contained OPM (Quspin model QZFM2 Zero-Field Magnetometer) placed at different positions in the region outside the coil planes.  Alternating magnetic fields of $\sim2$ \si{\nano\tesla} are applied to each biplanar coil at a fixed frequency in the range 10-20 \si{\hertz}.  The OPM-sensor output is sampled for 30 s and then Fourier transformed to separate the signals corresponding to each coil.  The signal amplitudes at each frequency are scaled to the response of a known test field, in order to convert each stray-field value into a percentage of the field at the coil center.

For the case of the oversize coils, the OPM is placed on a 3-axis translation stage and stray field along the $x$ axis is measured in point-by-point fashion on a 2.5 mm grid in the $x$-$y$ plane.  For the case of the miniature coils, the OPM position inside the magnetic shield is fixed while the coils were moved on a grid corresponding to 2-mm translation along and $45$\si{\degree} rotation about the $x$ axis.

\subsection{Single-beam OPM}
Circularly polarized light of around \SI{1}{\milli\watt} optical power at 795 nm is collimated, passed through a hole in the magnetic shield and then through the MEMS vapor cell.  The beam source is a VCSEL and the beam is conditioned using (in order) a Keplerian beam expander, a linear polarizer and a zero-order quarter-wave plate.  The transmitted beam is finally detected outside the shield using an amplified Si photodetector (Thorlabs model PDA36A2).

The laser wavelength for optical pumping depends on the pressure-shifted atomic resonance frequency and is tuned on resonance by minimizing the transmission of circularly polarized light when a large transverse dc field of the order of $10$ \si{\micro\tesla} is applied to the atoms.  The residual magnetic field is then nulled by applying direct current to the biplanar coils. A 3-channel, low-noise current source (Twinleaf CSB-10) is used to supply current to each coil through a \SI{500}{\ohm} shunt resistor. 

To null the background field, optical transmission measurements are performed around the zero-field Hanle resonance.  A slow (5 Hz) linear current ramp is applied to the X coil, to produce a field from $B_x = $ $–$\SI{250}{\nano\tesla} to +\SI{250}{\nano\tesla} and the zero-field resonance corresponding to maximum transmission of light through the vapor is resolved to a line width of a few tens of nanotesla.  The slow scan ensures that atomic polarization is in an approximately steady state during the sweep.  From the resonance curve, a ``sharpness'' figure of merit is calculated, equal to the maximum transmission amplitude divided by the full width at half maximum (FWHM) transmission.

Currents of the order of 2-3 mA are applied to the Y and Z coils ($I_y$ and $I_z$).  An automated grid search produces a three-dimensional plot of $(I_y,I_z, \mathcal{S})$ with a field resolution of \SI{100}{\pico\tesla} $\times$ \SI{100}{\pico\tesla}, thus giving the optimal nulling field as the point of highest sharpness.  The procedure is iterated one or two times more, starting from the best-guess zero field, but this step is usually unnecessary.

To perform field modulation and demodulation operations, a commercially available audio codec chip (Cirrus Logic CS4272) is interfaced with a low-cost microcontroller unit (ARM Cortex M7 model IMXRT1060, CPU speed 600 MHz, 64 bit math) via an Inter-IC Sound (I\textsuperscript{2}S) bus.  The outgoing I\textsuperscript{2}S signal from the microcontroller produces a sinusoidal voltage output of the codec (24 bit), to modulate the magnetic field produced by the X coil. The incoming I\textsuperscript{2}S signal is the ac-coupled signal, output from the amplified photodiode and digitized by the codec (24 bit).  The incoming signal is digitally demodulated against the modulation frequency on board the microcontroller, low-pass filtered using a double-cascaded second-order biquad filter and then the in-phase part of the signal (proportional to \autoref{eq:Pz2}) is sampled at 3 ksps and output to a PC. 

\section{Results}
\label{sec:results}

\subsection{Coil specification}
Two biplanar XYZ coil sets are presented in this work, as follows.

\subsubsection{Oversize coils}
A three-axis biplanar coil system is designed to be fabricated on a PCB and retrofitted to an existing laboratory OPM setup where miniaturization is not a primary concern.  Shown in \autoref{fig:coil photographs}(a), the coil system comprises two \SI{40}{mm} by \SI{40}{mm} planes printed as a two-layer PCB on a single sheet of flexible polyimide, which is cut as indicated and then folded in half to give a \SI{15}{mm} inter-plane distance.  The support structure for the PCB is a plastic oven cavity containing a MEMS vapor cell of \textsuperscript{87}Rb with approximate $xyz$ dimensions \SI{5}{mm} $\times$ \SI{5}{mm} $\times$ \SI{1.5}{mm}\cite{Karlen2017}; the coils are therefore an order of magnitude larger than the MEMS cell and described as ``oversize''. The coil planes include holes for screws and alignment rods as well as a large central hole in one of the planes for the vapor cell and heating element, which the PCB design has to accommodate.

The field specification for the oversize coils is as follows: (1) a homogeneous magnetic field over the volume occupied by the MEMS vapor cell, specified as $<$0.25\% error per field component in a spherical target region of radius 2 mm at the coil center, equidistant between the coil planes; (2) below 2\% stray field at the inner walls of a commercial magnetic shield in which the OPM is placed; and (3) a target field-to-current ratio for each coil of between 100 and 200 \si{\nano\tesla\per\milli\ampere} (see the design details given in \autoref{sec:methods}),  The final coil design includes coils for fields along the $x$ (X coil), $y$ (Y coil) and $z$ (Z coil) axes plus a Hemholtz coil along the $x$ axis (XH coil).

\subsubsection{Miniature coils}
Another 3-axis biplanar coil system is designed for an ultracompact self-shielded zero-field OPM package to be used in MEG studies.  For this design, small-form-factor coils are required to produce highly homogeneous magnetic fields at the vapor cell with minimal stray field outside the package.  Additionally, coil placement is secondary to the placement of other components in the OPM such as optics and thermal insulation.  The geometric specification is thus a system comprising two planes on either side of a cylindrical target region of diameter \SI{3.1}{\milli\meter} and thickness \SI{1.5}{\milli\meter}, corresponding to the MEMS atomic vapor cell \cite{Karlen2017}, and overall PCB footprint of \SI{17}{mm} $\times$ \SI{17}{mm} excluding connecting wires.  The field specification is as follows: (1) 0.5\% inhomogeneity per $x$, $y$ and $z$ field component within the target region; (2) a stray field, again for each field component, below 0.5\% of the target field at 20-mm distance from the homogeneous target region.  Orthogonality of the miniature coils is also well controlled. The computed mean angular distances between the three coils were in the range \SI{89.7}{\degree} to \SI{90.6}{\degree}.  The manufactured PCB and assembled coils are shown in \autoref{fig:coil photographs}(b) (see the design and optimization-process details given in \autoref{sec:methods}.

\subsection{Field-to-current ratio and homogeneity}
\label{sec:internalfields}

\begin{figure*}[t]
\centering
\includegraphics[width=16cm]{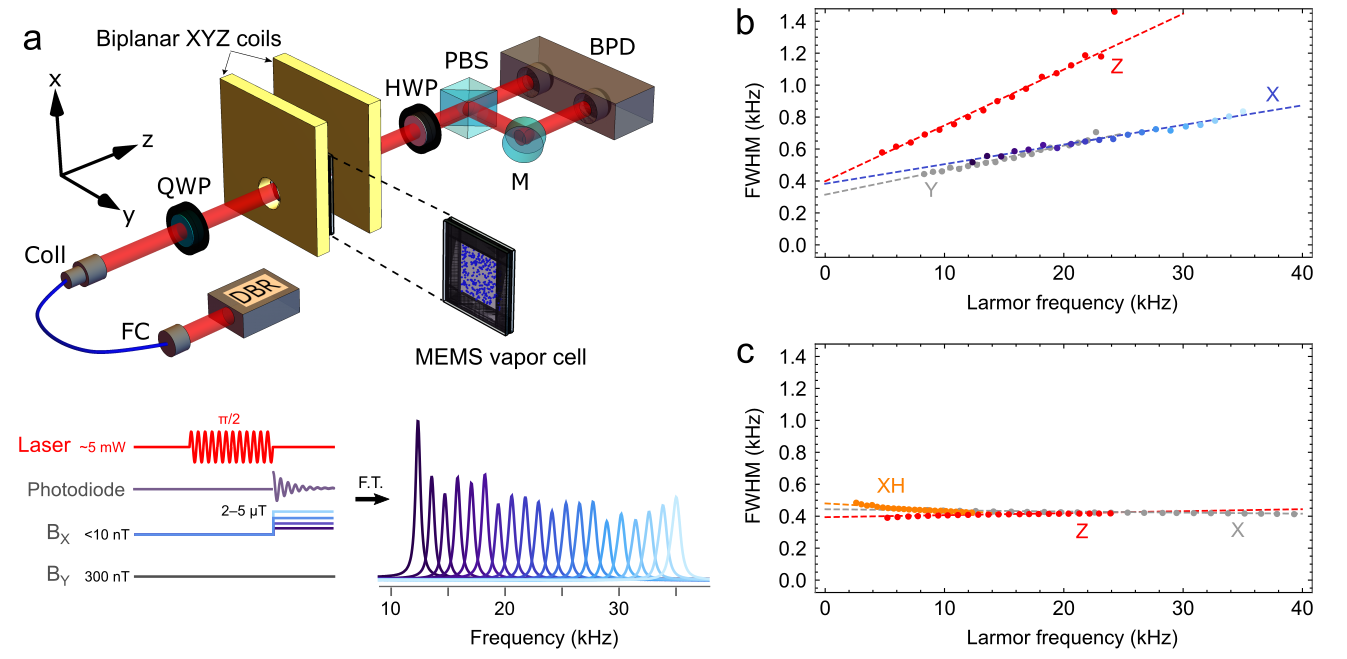}  
\caption{The free-induction spectroscopy of \textsuperscript{87}Rb in a MEMS vapor cell versus the local magnetic field supplied by biplanar coils: (a) The optical setup for measurement of atomic spin precession and transverse relaxation rates:  DBR, distributed Bragg reflector laser; FC, fiber collimator; QWP(HWP), 795-nm quarter-(half-)Wave plate; PBS, polarizing beamsplitter cube; M, mirror; BPD, balanced photodiode-pair detector; FT, Fourier transformation. (b) The spectral broadening of the magnetic resonance line versus atomic Larmor frequency for magnetic fields applied by each miniature coil, X (long axis, blue), Y (short axis, gray) and Z (laser beam axis, red).  The inhomogeneous broadening per unit magnetic field corresponds to the gradient of each best-fit straight line, which equals the ratio of the gradient to the applied magnetic field for the corresponding coil: $\kappa = |\frac{\delta B}{B_0}|$ (where $\kappa_X=1.2\%$, $\kappa_Y=1.5\%$ and $\kappa_Z=3.5\%$); (c) The field dependence of the transverse relaxation rate for the oversize coils. FWHM, full width at half maximum.}
\label{fig:FID signals}
\end{figure*}

The specification of magnetic field inside the biplanar coils is validated through atomic FID measurements.  As shown in \autoref{fig:FID signals}(a) and described in the \autoref{sec:methods}, a MEMS \textsuperscript{87}Rb vapor cell is placed at the intended location in the center of the coil assembly and illuminated with D\textsubscript{1} laser light to produce atomic polarization.  The atomic polarization is then allowed to precess in the magnetic field supplied by the coils and its magnetic moment along the beam axis is detected by Faraday rotation in the same laser beam.  The rotation signal is proportional to $\exp(-R_2^\ast t)(B_x,B_y,0)\cos(2\pi f_L t)/|\bm{B}|$.  In this expression, $\bm{B}$ is the average magnetic field over the volume where the beam intersects the atoms, $2\pi f_L \equiv \gamma |\bm{B}|$ is the atomic Larmor frequency and $R_2^\ast$ is the transverse relaxation rate.  
 
Measurements of Larmor frequency for a set of applied currents, such as those shown in the lower part of \autoref{fig:FID signals}(a) for the X coil, allow a precise estimation of the field-to-current ratio by fitting the derivative term in $B_\alpha = (\partial B_\alpha/\partial I_\alpha)I_\alpha$ with $\alpha \in \{ x,y,z \}$.  Experimental values of $\partial B_\alpha /\partial I_\alpha$ for both oversize and miniature coil systems are shown in \autoref{tab:coil_internal_results}.  Overall, there is good agreement between the experimental values and those predicted by the stream-function method during the design phase (also shown in \autoref{tab:coil_internal_results}), particularly for the miniature coil set, where deviations are below 1\%.  The deviation between experimental and predicted values for the oversize coil set are slightly higher at 2\%-4\%, which may be attributed to the design modification needed to accommodate all coil traces on the two-layer flexible PCB substrate.  

\begin{table*}
\centering

\begin{tabularx}{\textwidth}{lYYYYY}
 \hline \hline
 & Coil & \multicolumn{2}{c}{Field-to-current ratio (\si{\nano\tesla\per\milli\ampere})} & \multicolumn{2}{c}{Inhomogeneity (\%)}\\
&& Computed & Experimental & Computed & Experimental \\
 \hline

\multirow{3}{*}{Miniature} & X & 180.1 &    179.1         & 1.1 & 1.2          \\                  
& Y & 119.3 &   115.7        & 1.3 & 1.5            \\
& Z & 184.5 &    184.1        & 2.4 & 3.5  \\ \hline
\multirow{4}{*}{Oversize} & X &    194.37 & 192.6(3)     & 0.55 &   $<$0.1        \\
& Y & 112.43 &    116.3(2)          & 0.48 &  $<$0.1      \\
& Z & 112.38 &   116.26(16)         & 0.46 &     $<$0.1\\
& XH (Helmholtz) & 59.16 &    60.1(1)          & 0.01 &  $<$0.1  \\
 \hline \hline
\end{tabularx}
\caption{The computed and FID-measured internal magnetic field parameters for the oversize- and miniature-coil systems.  The computed field-to-current values refer to the sensitive volume of a MEMS vapor cell at the coil centers and are thus directly comparable to the FID values. The computed inhomogeneities are the 95th percentile of field deviation over the MEMS cell, while the experimental value is obtained from the slope of transverse relaxation rate versus Larmor frequency for each series of FIDs.}
\label{tab:coil_internal_results}
\end{table*}

The relaxation rate $R_2^\ast$ is obtained from the FWHM ($\Delta\nu$, in Hz) of the Fourier-transformed FID signal,  $R_2^\ast / \pi = \Delta\nu$, and provides direct information on the field inhomogeneity across the target region of the vapor cell.  We can write $R_2^\ast= R_2 + R'$, where $R_2$ is the homogeneous relaxation rate including spin-exchange, spin-destruction and optical-pumping relaxation effects, and $R'$ is the rate of inhomogeneous broadening, which is linearly proportional to the gradient.  To avoid compromising the OPM performance, gradients across the sensor atoms should be low enough to ensure $R' \ll R_2$.  Here, we quantify the field inhomogeneity through a derivative $\kappa_\alpha = \partial R_2^\ast /(\pi \gamma \partial B_\alpha)\equiv\partial\Delta\nu /\partial f_L$ for each coil, by assuming $\kappa_\alpha = |\Delta{B}_{0,\alpha}/B_{0,\alpha}|$, where $\Delta B_{0,\alpha}$ is the effective variation in magnetic field over the cell volume at a given $B_0$ along axis $\alpha$.  Values of $\Delta B_{0,\alpha}$ are determined in a similar way to the field-to-current ratios, by collecting a series of FID spectra for different applied currents and fitting the slope of $\Delta\nu$ vs.\ $f_L$.  

Experimental FID linewidths obtained using the miniature coil set are shown in  \autoref{fig:FID signals}(b) and similarly using the oversize coil set in \autoref{fig:FID signals}(c).  The miniature coil set yields values of $\kappa_{x} = 1.2\%$, $\kappa_{y} = 1.5\%$, and $\kappa_{z} = 3.5\%$, while for the oversize coil set, $\kappa_{x}$, $\kappa_{y}$ and $\kappa_{z}$ are all below $0.1\%$.  It is unsurprising to find that the miniature biplanar coils produce fields with lower homogeneity than the oversize counterparts.  The target volume of the miniature coils occupies a larger percentage of overall coil and therefore imposes a greater challenge for the design optimization.  Furthermore in this case study, higher priority is given during the design phase to the minimization of stray fields.  The smaller coil size is also more susceptible to errors in the interplane distance and plane alignment made during coil assembly.  However, despite these trade-offs, the inhomogeneity remains relatively small and within 10\% of the values expected for the design.  

The relaxation data in \autoref{fig:FID signals}(b) and \autoref{fig:FID signals}(c) can also suggest limits imposed on the OPM dynamic range by field gradients.  If the noise is limited by atomic spin projection noise, for example, the sensitivity will become significantly degraded above fields where $R'$ becomes on the order of $R_2$, or about 20 kHz Larmor frequency (\SI{30}{\micro\tesla} field) for the miniature X coil.  For the oversize coils, the limit should occur at much higher field, because there is no noticeable slope of the relaxation rate versus applied field.

\subsection{External field characterization}

\begin{figure*}
\centering
\includegraphics[width=15cm]{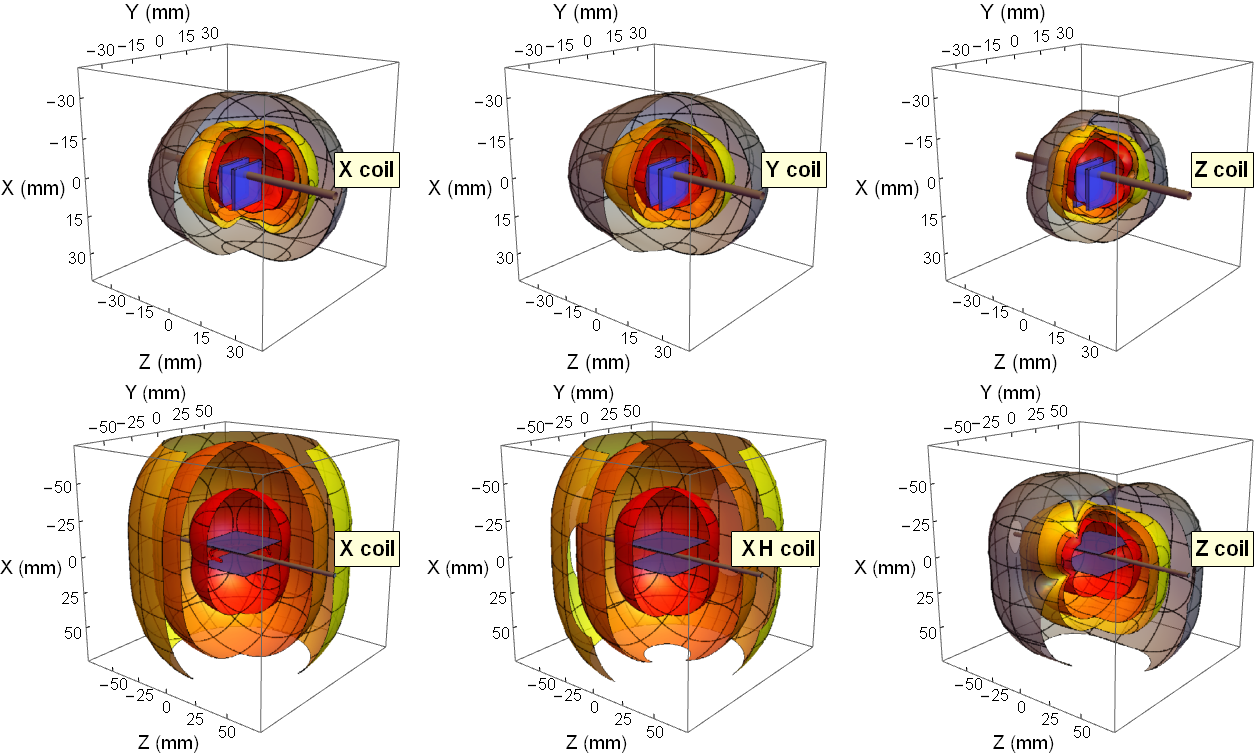}
\caption{Calculated stray-field maps for the miniature (top row) and the oversize (bottom row) biplanar coils.  The coil planes are indicated by blue shading and the intended path for a light beam is shown by the gray cylinder, which defines the $z$ axis.  The surfaces of increasing enclosed volume represent three-dimensional contours of decreasing total stray field; where visible within the plot range, the colors red, orange, yellow and gray represent 5\%, 1\%, 0.5\% and 0.1\% of the value of the field at the origin, respectively.  The mesh lines correspond to the tick marks on the $x$, $y$ and $z$ axes. The bottom row includes both the stream-function-optimized X coil and the Helmholtz coil XH. A plot for the oversize Y coil is not shown because after 90\si{\degree} rotation about $x$ it is equal to that for the Z coil.}
\label{fig:computed_stray_field}
\end{figure*}

The fringe field specifications of the two biplanar coils designs are also validated by comparing experimental data with the simulation results.  The ``bfieldtools'' software package provides a convenient route to calculate the field profile from the optimized current paths, using a single built-in function.  Three-dimensional representations of the stray field can then be plotted and inspected; for instance, the isosurface maps as plotted in \autoref{fig:computed_stray_field}, to show surfaces of constant stray-field magnitude.  The isosurfaces in \autoref{fig:computed_stray_field} indicate the stray field between 5\% (red) and 0.1\% (gray) of the field at the coil center.  The overall most compact isosurfaces occur for the miniature Z coil, where total field does not exceed 0.1\% at any point further than 30 mm from the coil center.   

In practical applications where it may be important to control external stray fields, usually only a small portion of the field map is relevant.  Examples would be known positions where neighboring sensors are to be placed, or of predicted field hotspots.  In these cases, the task of comparing experimental versus calculated fields then becomes a lower burden (especially where multiple prototypes are to be tested) because experimental field mapping needs only be performed around the regions of practical concern.  For example, in MEG, OPM sensors are typically arrayed on the head surface to measure the magnetic field component normal to the scalp\cite{iivanainen_measuring_2017}.  If the magnetic field at each OPM sensor were supplied by the miniature coil set from \autoref{fig:coil photographs}(b) then $x$ would be the field-sensitive axis and the nearest-neighbor sensors would lie in the $y$-$z$ plane. 

\begin{figure*}
\centering
\includegraphics[width=0.8\textwidth]{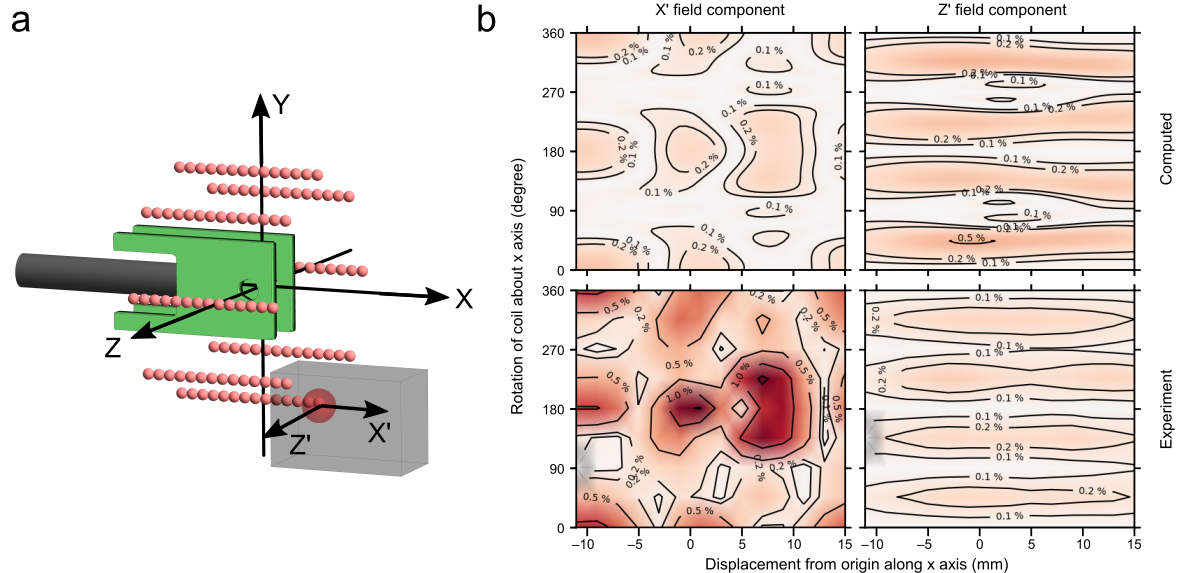}
\caption{Stray-field maps for the miniature X coil. (a) The experimental setup.  Pink spheres indicate the location of the experimental data points in (b). (b) computed and empirically measured magnitude of the stray-field components of the X coil along two axes tangential to the surface $\sqrt{y^2+z^2} = 20$ mm.}
\label{fig:Field_Mapping_miniature}
\vspace{0.7cm}
\includegraphics[width=0.8\textwidth]{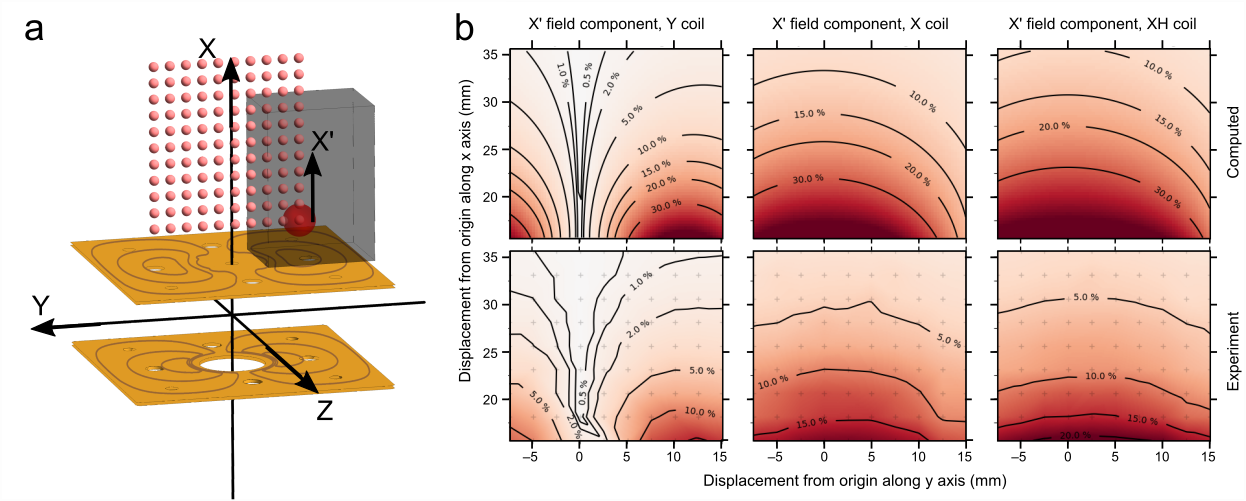}
\caption[Coil Field Mapping]{Stray-field maps of X,Y,Z and XH oversize coils: (a) The experimental setup, showing the relative arrangement of the PCB planes and the magnetometer used to sample the $x$ stray-field component in the $x$-$y$ plane. The current loops for the biplanar Y coil are plotted in brown. (b) The computed and experimentally measured magnitude of the stray-field component along $x$.  The pink spheres in (a) indicate the location of experimental data points shown by crosses in (b)}
\label{fig:Field_Mapping}
\end{figure*}

In this work, the fringe field of the miniature coils is validated on a cylindrical surface with radius $(y^2 + z^2)^{1/2}$ = 20 mm and principal axis along $x$.  The chosen value of 20 mm corresponds to a proposed minimum approach distance between vapor cells in a MEG array.  As shown in \autoref{fig:Field_Mapping_miniature}(a) (see also \autoref{sec:methods}), specific components are experimentally mapped in point-by-point style by translating and/or rotating the coils with respect to an auxiliary OPM sensor (QuSpin model QZFM Zero-Field Magnetometer) located in the fringe field.  The orientation of the magnetometer is such that field components tangential to the cylinder surface are measured and the results are plotted in \autoref{fig:Field_Mapping_miniature}(b).

Experimentally, the shape of the stray field contours agrees well with the ``bfieldtools'' simulations for all coils (shown in \autoref{fig:computed_stray_field}), and for the Y and Z coils the strength of the stray field also agrees quantitatively.  For some unknown reason, the stray fields of the X coil along the $x'$ direction turn out to be much higher than expected, with a hot spot at the position $x=7$ mm (\SI{235}{\degree}) of around 1.2\% experimental versus 0.3\% computed.  A partial reason for this result may be a small error in the inter-plane separation; coils misaligned by only 0.5 mm along the plane normal direction cause the computed stray field to reach around 0.5\% at the 20-mm sensor standoff distance.  In practice, the actual standoff distance may be less than the expected 20 mm due to imprecise sensor positioning and/or the finite width of the QZFM vapor cell\cite{quspinQZFMGen3} (approximately \SI{1.5}{\milli\meter} diameter).  Moreover, in practice, the QZFM sensitive axis is not exactly parallel to $x'$ and additionally picks up around 20\% of a large stray-field component along the direction $y'$.  If the standoff distance is reduced to 18 mm, then the computed total field at the hot-spot location becomes the order of 1\% and therefore the experimental value of 1.2\% could be explained by combined errors in coil alignment and sensor standoff of around \SI{1}{\milli\meter}.  Aside from the issue of alignment tolerance, the specific defect described here is highly localized, likely because the coil set is designed for minimal stray field on a spherical surface rather than on a cylinder and in some applications may not pose major problems.  For example, it is possible to array the miniature coils on a square grid in the $y$-$z$ plane at the \SI{90}{\degree} and \SI{270}{\degree} locations, which avoids the worst hot spots.  In future, the design-optimization objective could be defined with a stronger stray-field constraint, including stray fields on  multiple surfaces to take tolerances into account and/or avoid local hot spots.

A validation of the stray field of the oversize coils is presented in \autoref{fig:Field_Mapping}.  The contour plots show a field map generated using a procedure similar to that described above for the miniature coils, but only the $x$ component of the field is studied.  The experimental stray fields of the X and Y coils along $x$ are somewhat smaller than the corresponding computed results, by a factor of around 2.  The field of the Z coil (not shown) along $X'$ is roughly 10 times weaker than that of the X coil, measuring around 1\%-2\% near to the coil plane and less than 0.1\% upon moving \SI{15}{mm} away from the plane, along $x$ -- which is expected, since the sensor axis is orthogonal to the principal axis of the coil.  The stray-field map for the Y coil also reveals the symmetry plane along which the stray field is negligible ($<$ 0.1\%).

\autoref{fig:Field_Mapping}(b) also allows comparison between the fringe fields of the XH and X coils.  The two maps are extremely similar to one another, which is in agreement with the computational prediction.  The result is interesting because the single-loop XH coil takes up significantly less space on the PCB than the optimized X coil.  In future, therefore, a smaller triple-coil set could be easily made on a two-layer flexible PCB, by using only the optimized Y and Z coils plus the XH coil.

\subsection{Open-loop OPM performance}

In each of the coil designs presented, the central homogeneous-field target region corresponds to the intended location of optically pumped atoms in a MEMS vapor cell.  To provide a more complete characterization of the biplanar coil performance, we demonstrate a simple application of the system as a magnetometer.  

\begin{figure*}
\centering
\includegraphics[width=0.7\textwidth]{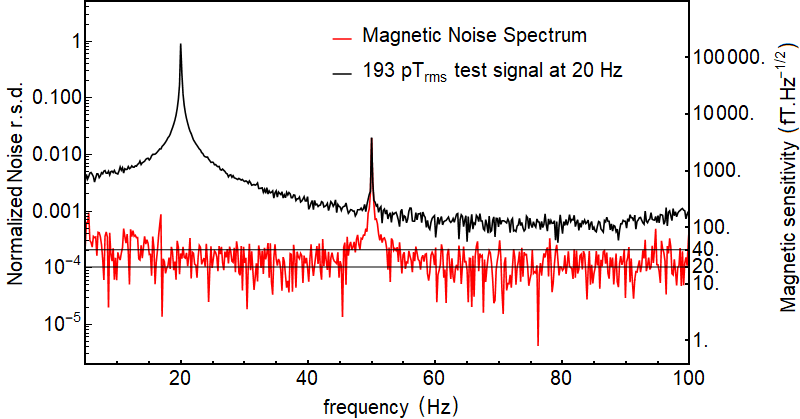}   
\caption[Sensitivity]{A demonstration of zero-field-resonance magnetometry: open-loop magnetic noise spectra obtained using a MEMS \textsuperscript{87}Rb vapor cell contained within the oven assembly of \autoref{fig:coil photographs}(a). A 533-Hz, 24-nT modulation field is applied using the oversize X coil. The spectra are calculated as the root-mean power spectral density of surplus optical-pumping light detected after the vapor cell and after demodulation, which we scale to a response curve for ac test signals of known amplitude in the range 1 Hz to 200 Hz.  The red curve indicates the noise background.  The black curve is the response to a 20-Hz, $193$-pT$_{\mathrm{rms}}$ sinusoidal test signal applied in the X coil in addition to the modulation field.}
\label{fig: Sensitivity}
\end{figure*}

We focus on a zero-field-resonance magnetometer comprising the MEMS vapor cell, the oven and the oversize PCB coils (\autoref{fig:coil photographs}(a)).  This assembly is placed inside a four-layer Mu-metal magnetic shield, while the laser source and other optical components to generate circularly polarized light are outside the shield and fixed on an optical table, as detailed in \autoref{sec:methods}.  The vapor-cell dimensions are identical to those used in the FID measurements presented in \autoref{sec:internalfields}.  The vapor cell composition, however, is different, with the \textsuperscript{87}Rb vapor density and the N\textsubscript{2} buffer gas pressure optimized to give the largest response of the quadrature signal (\autoref{eq:Pz2}) per unit field component $B_x$.  The specific cell used has a zero-field-resonance line width of $B_1$ = \SI{35}{\nano\tesla} and the best quadrature signal, empirically speaking, is obtained for modulation parameters $B_{\rm mod}=$  \SI{24}{\nano\tesla} and $\omega_{\rm mod}/2\pi = 533$ Hz.  The XH coil is used to provide the modulation field and the X, Y and Z coils are used to null the total background field from several tens of nT to below 5 nT.  It is observed, as shown in \autoref{fig: Sensitivity}, that the above settings can produce a demodulation signal with a field-equivalent baseline noise in the range of \SI{20}{\femto\tesla\per\sqrt\hertz} at frequencies 20--100 Hz.  The obtained sensitivity is similar to that of commercial-prototype OPMs that use miniature glass-cuboid\cite{osborne_fully_integrated_2018} or etched-silicon MEMS\cite{alem_opticsexress2017,nardelli_conformal_2020,twinleaf_microserf} cells to contain the alkali medium, which reach on the order of 10--30 \si{\femto\tesla\per\sqrt\hertz}.

With regard to the noise floor of \autoref{fig: Sensitivity}, it is estimated that the thermal noise of the magnetic shield contributes around \SI{10}{\femto\tesla\per\sqrt\hertz}. System electronics excluding the VCSEL driver contribute an effective field noise of less than \SI{2}{\femto\tesla\per\sqrt\hertz}.  The VCSEL driver and other optical-noise sources are attributed to the remainder of the noise budget.  A broader discussion of the MEMS magnetometer sensitivity, including results for the OPM in closed-loop mode and the miniature coils setup of \autoref{fig:coil photographs}(b), will be presented separately in an upcoming publication.

\section{Discussion and conclusion}

Coil design is a multifaceted engineering task in which a large number of competing factors must be taken into account and balanced, including size, geometry, field gradient, stray field, manufacturability and cost.  Classical coil designs such as Helmholtz coils can provide good performance, but are not flexible in terms of their geometry and have a significant stray field.  Sophisticated methods such as the stream-function coil-design framework applied in this work are required in applications with demanding geometrical and/or electrical constraints. 

The miniature three-axis coil set designed in this work is intended to be positioned around an individual atomic vapor cell in an OPM array.  The design specification of the OPM introduces significant constraints on the field profile both inside and outside the coil to minimize inhomogeneous broadening of the magnetic resonance line of the atoms and mitigate crosstalk fields, respectively.  Given that planar PCB manufacturing is highly developed, scalable and precise, the stream-function optimization framework applied on planar surfaces is especially suited to this design task, since the geometrical limitations of a planar structure can be compensated for by intricate in-plane current paths.  Here, the specification of the miniature coils is experimentally validated as already close to the MEG design goal, even taking into account potential sources of error encountered experimentally, such as in the stray field.

In the future, improved planar coils could be made with additional design optimization.  The effect of stray fields could be reduced by further scaling down coil size.  For instance, the magnetic field at the center of a Helmholtz coil (defined by the loop radius and the inter-plane distance $R$) is proportional to $1/R$, while at displacements further than $\pm R$ along the main symmetry axis, the magnetic field is proportional to $R^2$.  The scaling between the stray and the center field therefore goes as $R^3$ and coils printed directly onto the window substrate of the vapor cell may be an option\cite{Edri2021}, where permitted by cell geometry and aspect ratio.  The Helmholtz coil is already close to optimal with regard to magnetic gradients because, by definition, the first non-zero derivative along the main axis ($\alpha$) of the coil is ${\rm d}^4B_\alpha/{\rm d}\alpha^4$.  However, for a typical MEMS atomic vapor cell where the beam direction $z$ is perpendicular to the plane of the windows, a Helmholtz coil ($\alpha=z$) cannot provide the transverse modulation field ($x$) and therefore high-performance X- and Y-coil geometries must also be found.  On the basis of PCB trace width and intertrace spacing alone, the coils presented in this work may be shrunk by a factor of 2-3 and still be manufactured using currently available PCB techniques (0.002 inch $\approx$ \SI{50}{\micro\meter} minimum feature size); thus the biplanar design approach can, in principle, achieve lower stray fields by one order of magnitude.  Potentially, such coils may address the secondary problem of crosstalk when the modulation fields of adjacent sensors is not exactly in phase, where the stray-field constraint is even tighter.

A smaller interplane spacing in the case of biplanar current surfaces would also help reduce the magnitude of fringe fields because the solid angle covered by the planes (as seen from the target region) can be increased.  Another method would be to add one or more additional current layers, the main main purpose of which is to create a self-shielded coil\cite{Wu2020}, e.g., a tetraplanar coil.

Finally, an alternative to the zero-field-operation requirement may be given by total-field OPMs that also reach high sensitivities \cite{Sheng2013,https://doi.org/10.48550/arxiv.2112.09004} and can be miniaturized \cite{Schwindt2004} but implement complex optical configurations such as dual or multipass beams\cite{Lucivero2021} and are partially limited by spin-exchange relaxation.  An all-optical total-field OPM designed to eliminate sensor crosstalk has been demonstrated for MEG in an ambient (unshielded) environment\cite{Limes2020} with sensitivity reaching \SI{15}{\femto\tesla\per\sqrt\hertz} in gradiometer mode (\SI{10}{\pico\tesla\per\sqrt\hertz} single-sensor sensitivity). The utility of such OPMs in sensor arrays for MEG applications remains under study\cite{Clancy2021}.

\section*{Acknowledgement}
The work was funded by:
the European Union’s Horizon 2020 research and innovation programme under project macQsimal (Grant Agreement No.\ 820393);
Horizon H2020 Marie Sk{\l}odowska-Curie Actions projects ITN ZULF-NMR  (Grant Agreement No.\ 766402) and PROBIST (Grant Agreement No.\ 754510); 
the Spanish MINECO project OCARINA (the PGC2018-097056-B-I00 project funded by MCIN/ AEI /10.13039/501100011033/ FEDER, “A way to make Europe”); 
the Severo Ochoa program (Grant No.\ SEV-2015-0522); 
the Generalitat de Catalunya through the CERCA program; 
the Ag\`{e}ncia de Gesti\'{o} d'Ajuts Universitaris i de Recerca under Grant No.\ 2017-SGR-1354;  the Secretaria d'Universitats i Recerca del Departament d'Empresa i Coneixement de la Generalitat de Catalunya, cofunded by the European Union Regional Development Fund within the ERDF Operational Program of Catalunya (project QuantumCat, ref.\ 001-P-001644);
the Fundaci\'{o} Privada Cellex; 
and the Fundaci\'{o} Mir-Puig; 
M.C.D.T.\ acknowledges financial support through the Junior Leader Postdoctoral Fellowship Programme from the ``La Caixa'' Banking Foundation (project LCF/BQ/PI19/11690021).  
We also thank Jacques Haesler, Sylvain Karlen and Thomas Overstolz of the Centre Suisse d'Electronique et de Microtechnique SA (CSEM) in Neuch{\^a}tel (Switzerland) for supplying the MEMS vapor cells.

\bibliography{references}

\end{document}